# ARTICLE

# Vector-free DNA transfection by nuclear envelope mechanoporation

Leyla Akh,[a,1] Apresio K. Fajrial,[b,1] Benjamin Seelbinder,[b] Xin Xu,[b] Wei Tan,[ab] Jill Slansky,[c] Corey P. Neu,[abd] Xiaoyun Ding.[abd*]

Genetic engineering of cells has a range of applications in treating incurable diseases. Plasmid DNA is a popular choice of nucleic acid for cell engineering due to its low cost and stability. However, plasmid DNA must survive the protective mechanisms present in the cell's cytoplasm to enter the nucleus for translation. Many of the existing methods for nucleic acid delivery, such as chemical-based and virus-based delivery, suffer from drawbacks induced by the nucleic acid carrier itself. Mechanical methods present an alternative to nucleic acid carriers by physically producing openings in the cell to deliver cargos. However, in most systems, the cell membrane openings are too small to deliver large cargos, or the poration process leads to low cell viability. In this study, we present a microfluidic device with integrated high aspect ratio nanostructures that repeatably rupture the cell membrane and nuclear envelope. These sharp-tipped nanolancets penetrate the cell deep enough to allow direct delivery of cargos into the nucleus, but still allow for cell recovery after treatment. We show the device's ability to deliver cargo to a variety of cell types while maintaining high viability. Then, we demonstrate the rapid onset of plasmid DNA expression that results from direct nuclear delivery of naked DNA, showing expression speeds comparable to microinjection, but with significantly greater throughput. We envision the use of this device as a tool to quickly produce high quantities of genetically engineered cells to treat a myriad of diseases.

## Introduction

Cell transfection, the process of inserting genetic material into cells to modify their behavior, has a wide variety of applications in medicine, ranging from cancer treatment to therapies for genetic disorders[1]. The most common method for commercial production of genetically modified cells uses viral vectors, where viruses are modified to carry nucleic acid sequences and deliver them to host cells for expression[1,2]. However, viral-vector-based methods suffer from a number of drawbacks, including high manufacturing costs and potential immunogenicity to patients[3]. Current efforts in cell engineering focus on alternative methods to engineer cells at a lower cost, higher throughput, and without the need for viral vectors[3].

Non-viral methods for cargo delivery to cells have emerged to overcome the limitations of viral delivery, including chemical-based and membrane-disruption-based approaches[4]. In chemical-based delivery methods, nucleic acids complex with chemical reagents to form particles capable of entering the cell[2]. The use of chemical transfection methods requires a compromise between transfection reagent-induced cytotoxicity and sufficient transfection efficiency[1]. In membrane disruption methods, the cell membrane is penetrated to allow for the entry of cargos, and the cell must then undergo active membrane repair processes[2]. Examples include: microinjection, where cargos are injected into individual cells, yielding a high transfection efficiency but low throughput; electroporation, where the application of electrical voltage opens transient micropores in the plasma membrane to allow cargo entry, a process that may irreversibly damage cells or lead to cell death; and other methods such as magnetofection and ultrasound- and laser-assisted transfection, where openings in the cell membrane are induced to allow cargo entry, which suffer from similar drawbacks as electroporation[1,2]. Methods have also emerged that utilize cell squeezing or fluid shear to induce pores in the plasma membrane[5–8], but these methods are typically not capable of direct nuclear delivery. Another method uses penetrating needles in capture sites that draw cells to the needle by aspiration. This platform has potential to deliver cargos directly to the nucleus, but suffers from low viability[5]. A publication by Man et al. uses sharp metal tips to concentration heat from laser irradiation to generate microbubbles that disrupt the cell membrane. In this system, eight sharp tips contact each cell, producing more damage than systems that generate only one pore[9].

Additionally, systems that incorporate nanostructures and electroporation together have emerged to balance the drawbacks of both approaches. These nanoporation technologies typically involve cell culture on an array of nanoneedles and voltage application through the

a. *Biomedical Engineering Program, University of Colorado, Boulder, CO 80309, USA.*
b. *Paul M. Rady Department of Mechanical Engineering, University of Colorado, Boulder, CO 80309, USA.*
c. *University of Colorado School of Medicine, CO 80309, USA.*
d. *BioFrontiers Institute, University of Colorado, Boulder, CO 80309, USA.*
* Corresponding author: Xiaoyun Ding, xiaoyun.ding@colorado.edu
[1] The authors contributed equally to this work

needles[6,10,11,11]. These systems use a lower voltage than bulk electroporation systems, reducing cellular damage and specifically targeting the plasma membrane pore, and they show delivery to a variety of cells. These methods suffer from throughput drawbacks, as samples are treated in batch processes as opposed to semi-batch or continuous flow processes which are capable of larger throughput. Typically, nanoporation systems also require intricate fabrication processes, pre-loading with cargo prior to cell treatment, and cell incubation on the nanoneedle array, all of which slow the production of engineered cells. Additionally, these methods do not specifically target the nucleus, although the system described Shokouhi et al. may incidentally allow for enhanced nuclear delivery of cargos. Although many high aspect ratio nanostructures have been previously reported as transfection tools, very few produce repeatable, high-throughput wounds on cells that puncture both the cell membrane and nuclear envelope[6].

Some cargos, such as DNA, require delivery to the nucleus of the cell to be translated into protein. To produce stable transfection in non-viral systems, DNA must be delivered to the nucleus of the cell to integrate into the host's genome[1,12]. Of the existing non-viral delivery mechanisms, few are capable of directly targeting or porating the cell's nuclear envelope, and instead rely on DNA integration into the nucleus during cell division[13]. This limitation means that cells take hours or days to express transfected genes, increasing the manufacturing time and cost for these therapies.

Microfluidics have merged as a platform for precise manipulation and control of cells and fluids[14–16]. In this study, we report a new microfluidic technology capable of porating the nuclear envelope through mechanical piercing in high throughput. It consists of monolithically integrated nanostructures within microfluidic channels. Each constriction channel allows one cell to pass through at a time, and the sharp-tip nanostructure within each constriction pierces the cell membrane and nuclear envelope in high throughput. We first evaluate the performance of the device in dextran delivery studies, where we demonstrate plasma membrane and nuclear envelope poration via direct visualization with confocal laser scanning microscopy. Then, we show high rates of intracellular delivery of molecules of various sizes across multiple cell types, along with high cell viability results. Then, we show the device's capability for transfection of naked plasmid DNA and the ensuing rapid protein expression, highlighting nuclear envelope poration for immediate diffusive delivery of cargo in the nucleus. Finally, we validate the proliferative and genomic integrity of the cells to confirm that device treatment nor the cellular repair process fundamentally alter cell physiology. Our results give a new perspective on vector-free transfection for rapid and high-throughput production of engineering cells.

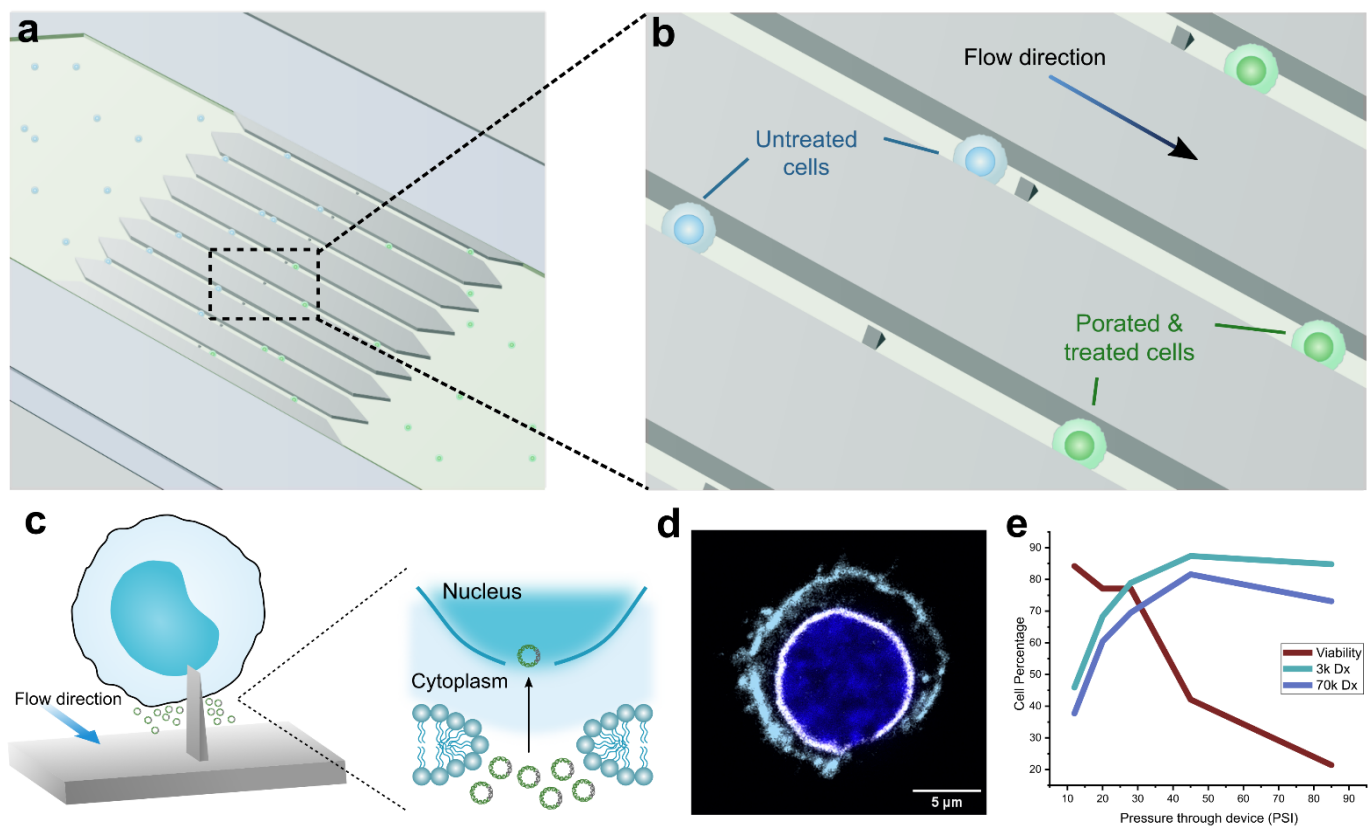

Figure 1: **(a)** Schematic showing typical NEST device operation, where cells in suspension flow through an array of microchannels. **(b)** An enlargement of panel (a), where suspended cells are shown to flow over nanolancets, which create pores for cargo delivery into the cell cytosol and nucleus. **(c)** Schematic showing the mechanism of cell poration by the nanolancets, which penetrate both the cell membrane and the nucleus. **(d)** confocal micrograph showing a HeLa cell after NEST device treatment. The cell membrane (light blue) shows a discontinuity that aligns with the discontinuity in the nuclear envelope (white), exposing the nucleoplasm (dark blue) to the extracellular space. **(e)** Cell viability and dextran delivery as a function of the flow pressure applied to the device. At lower pressures, viability is high but delivery of 3 kDa and 70 kDa dextran is low. As pressure increases, viability decreases, but the delivery efficiency of dextran increases.

## Results

**Design and characterization of the nanostructured microfluidic device.** High-aspect ratio and sharp tip nanostructures are an effective strategy for creating cell membrane openings, as reported in previous studies[14–16]. We developed the Nano-Engineered Surface Technology (NEST) device, which integrates sharp nanostructures into a continuous-flow microfluidic system (Fig. 1a-b). As cells flow through the microfluidic channels, they pass over the nanolancets that protrude from the bottom surface of the microchannel. These nanolancets are fabricated to be 2 µm in height and 4 µm long to precisely porate the plasma membrane and nuclear envelope. The sharp tip of the lancet measures approximately 100nm in width, and the angle of the tip is approximately 40°. These dimensions enable cargo delivery to the nucleus while preserving cell viability (Fig. 1c). For each cell type tested, the microchannel dimensions reflect the approximate diameter of the cell in suspension, ensuring contact with the nanolancet (see Methods and Supplementary Figure 1).

**Confocal microscopy confirms pores in the cell membrane and nuclear envelope.** To demonstrate poration of both the cell membrane and nuclear envelope, we performed confocal laser scanning microscopy. HeLa cells were stained to show the nuclear envelope and cell membrane, then flowed through the NEST device using a syringe pump,[17] then fixed and finally stained to show the nuclear envelope. Confocal imaging revealed poration of both the cell membrane and nuclear envelope (Fig. 1d), as evidenced by an absence of fluorescent signal in a portion of the nuclear envelope. In addition, there is an absence of fluorescent signal in a portion of the cell membrane coinciding with the location of the discontinuity in the nuclear envelope. Taken together, these observations suggest pore generation in both the plasma membrane and nuclear envelope via puncture with the sharp-tip nanostructure. The average diameter of the cell membrane pore was 7.4 µm (SD = 1.8 µm, n = 59), and the average diameter of the nuclear envelope pore was 2.6 µm (SD = 0.6 µm, n = 43).

**Pore generation and viability and a function of flow pressure.** The pressure of the fluid flow through the device can drastically impact pore formation and viability of the cells. To optimize the flow pressure, we delivered small (3kDa) and large (70kDa) fluorescently labeled dextran to HeLa cells under varying flow conditions in a device with a channel height of 12 µm and a width of 8 µm at the location of the nanoneedle (Fig. 1e). To optimize delivery and viability in HeLa cells, we selected a device of these dimensions with a blade length of 4 µm (see Supplementary Figure 2).  The cells were mixed with fluorescently-labeled dextran solutions prior to NEST treatment, then loaded into a syringe and flowed into the device. Predictably, a higher flow pressure reduces cell viability but increases the percentage of viable cells to which dextran is delivered. Selection of the optimum flow pressure for each cell type is a trade-off between viability and delivery; In HeLa cells, we selected the optimum pressure as 28 psi, where the viability is 77%, and 3kDa and 70kDa show 79% and 70% delivery, respectively (Fig. 1e).

At a flow pressure of 28 psi, a sample of 200,000 cells can pass through the device in under two seconds. The device design accommodates 100 individual nanolancet-containing channels, so cell clogs do not substantially hinder the flow rate of fluid through the device. Higher cell densities and larger sample volumes can lead to device clogging; however, large quantities of cells can be passed through the device in smaller batches, with breaks to clear clogs using water in backflow. More substantive clogs can be cleared with a mild bleach solution or an enzymatic cleaner to break down cellular debris.

**Fluorescent dextran imaging shows nuclear delivery of large molecules.** To demonstrate the nuclear envelope poration capabilities of the device, we delivered 70 kDa fluorescein-dextran molecules to HeLa cells. After NEST treatment, the cells were allowed to recover in medium for 15 minutes, then fixed in 2% PFA, washed, and analyzed by confocal microscopy. A control sample, where cells were incubated with 70 kDa fluorescein-dextran, was used to account for possible cell surface attachment of dextran.

The results in Fig. 2a show successful delivery of 70 kDa dextran to both the cell cytoplasm and the nucleoplasm, indicating not only poration of the cell membrane and nuclear envelope, but also the diffusion of molecules into the cell through these pores. The hydrodynamic radius of 70 kDa dextran exceeds the limit for passive permeability into the nucleus of HeLa cells;[18,19] therefore, the presence of dextran in the nucleus demonstrates transient nuclear envelope poration and the nuclear delivery capabilities of the NEST device. The delivery of large dextran molecules into the nucleus highlights the potential of this technology to be used for the delivery of nucleic acid directly into the nucleus.

**Intracellular cargo delivery with NEST device.** The results in Fig. 2b show that the NEST device can deliver 3 kDa (small) fluorescently-labeled dextran molecules to a variety of cell types with high efficiency: CT26, HeLa, A20 Jurkat, and primary human T-cells. Across the board, the device was capable of high delivery, up to 91% in CT26 cells, and a minimum of 58% achieved in primary human T-cells. Interestingly, in CT26 cells, the delivery efficiency of larger dextrans (70 kDa and 2000 kDa) were comparable to each other, but lower than the delivery efficiency of 3 kDa dextran (Fig 2c). These results emphasize that intracellular delivery mediated by the NEST device is less dependent on the cargo's molecular size compared to typical diffusion-based methods[20–22]. The primary factor that may contribute to the improved delivery of large cargos is the size of the pore produced by the NEST device; with cell membrane and nuclear envelope pores well exceeding the hydrodynamic radius of 2000 kDa dextran in water (15 nm [23]), the difference in delivery efficiency of medium-sized cargos, such as 70 kDa dextran, and large-sized cargos, such as 2000 kDa dextran, is negligible. However, smaller cargos, such as 3 kDa dextran, diffuse faster than larger ones,[24] explaining the difference in delivery between 3 kDa dextran and 70 and 2000 kDa dextrans. We interpret the delivery efficiency of dextran to represent the percentage of cells whose plasma membranes are damaged by the NEST device and that subsequently recover. Although nuclear delivery occurs during dextran delivery (Fig. 2a), the flow cytometric data in Figure 2b-c does not differentiate between nuclear and cytoplasmic delivery. Therefore, the data in Figure 2b-c is representative of the percentage of the cell population with plasma membrane pores from NEST treatment.

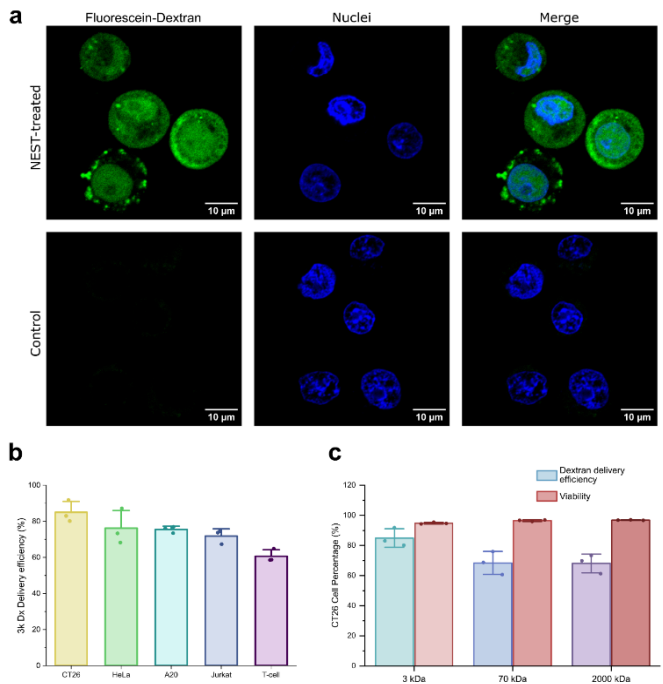

Figure 2: **Dextran delivery to multiple cell types. (a)** Confocal micrographs of cells with NEST treatment to delivery 70 kDa dextran (top row) or a sample that was co-incubated with 70 kDa but not treated with the NEST device. The top row of images show fluorescently-labeled dextran delivery to both the cytoplasm and nucleus as a result of NEST treatment, and negligible fluorescein-dextran signal in the untreated control sample. **(b)** Delivery of 3 kDa dextran into a variety of cell types, showing delivery exceeding 60% in all cell types tested. Delivery efficiency was assessed by flow cytometry immediately following NEST treatment, with each datapoint representing a sample of 10,000 cells. The error bars indicate s.d. (n = 3). **(c)** Delivery of fluorescently-labeled dextran molecules with a range of molecular weights into CT26 cells using the NEST device. The delivery efficiency and cell viability were assessed via flow cytometry immediately after NEST treatment, with each datapoint representing a sample of 10,000 cells. The error bars indicate s.d. (n = 3).

**Rapid protein expression from plasmid DNA transfection.** In the context of gene transfection, successful cytoplasmic delivery of plasmid DNA into the cell does not guarantee protein expression[25,26]. To be transcribed and translated into functional proteins, the transfected DNA must reach the nucleoplasm through a complex and highly regulated process of nucleocytoplasmic transport[27,28]. Exogenous DNA is also prone to degradation in the cytoplasm[29,30]; consequently, transfection of plasmid DNA by disrupting the plasma membrane alone is generally insufficient[21,22,31,32]. Chemical vectors can help protect genetic materials from cytoplasmic degradation and enhance nucleocytoplasmic transport, but the nuclear import process is slow and often requires nuclear envelope disruption via cell mitosis to enable cargo molecules to enter the nucleus[28].

We demonstrated that the NEST device can transfect plasmid DNA into HeLa cells with over 60% efficiency via direct nuclear envelope poration (Fig. 3a). As a comparison, we measured the transfection efficiency of a chemical vector, polyethylenimine (PEI), and a NEST device of identical dimensions fabricated without nanolancets. During the 24 hours of cell culture after delivery, PEI transfection achieved 48% efficiency, while the NEST device without nanolancets did not cause substantial transfection of the plasmid DNA. These results align with previously reported attempts at transfection of naked plasmid DNA by flowing cells into microconstrictions,[20–22] indicating the importance of the nanolancet in facilitating nucleic acid delivery in the NEST device.

Recent findings emphasize that the intact nuclear envelope blocks exogenous DNA from entering the nucleus, and transient nuclear envelope poration is necessary for protein expression from plasmid DNA, especially for cells that have not undergone mitosis[28,32]. We sought to demonstrate that the NEST device facilitates this transient nuclear envelope poration, leading to successful vector-free DNA transfection. We analyzed the dynamic expression of transfected DNA and compared the performance of the NEST device to chemical-based transfection methods. Fig. 3b shows the time course of pDNA expression in HeLa cells after transfection with enhanced green fluorescent protein (EGFP)-encoding plasmid DNA (pDNA) by various transfection methods: co-incubation with naked pDNA (indicated as "control" in Figure 3b), chemical-based transfection with PEI, and NEST treatment at different time points. The dynamics of fluorescence expression show an EGFP signal as early as 1h after NEST treatment, while PEI transfection shows no substantial EGFP signal before 12 h, and the co-incubation control fails to show any EGFP signal before 24 h. These results are corroborated by fluorescent microscopy of the transfected cells at various timepoints, shown in Supplementary Figure 3.

To better quantify the protein expression dynamics of the NEST device (Fig. 3c), we compare the performance of the device against PEI transfection, lipid-based chemical transfection (Lipofectamine 2000 ®, "Lipo"), and electroporation ("EP") from a previous report under identical cell culture conditions.[31] In both chemical transfection methods, the onset of protein expression occurred at 12 h, with no detectable protein expression before this time point. Electroporation showed modest protein expression of under 30% at the 12 h time point. In contrast, NEST transfection generated the most EGFP-expressing cells, nearly 50%, in the span of 4 hours. Moreover, the EGFP expression in NEST-transfected cells had a more uniform distribution and higher intensity compared to PEI-transfected cells (Supplementary Figure 4, Supplementary Figure 5). This key difference in expression time suggests nuclear poration via NEST treatment. For pDNA to be transcribed into RNA and translated into protein, it must enter the nucleus for transcription to occur. In typical transfection methods that do not target the nucleus, cell division is required to allow plasmid entry into the nucleus [28,32,33]; in those methods, expression occurs on time scales that coincide with the cell division timeline – in this case, 12-24 hours. The rapid expression shown by using the NEST device indicates that cell division is not necessary for nuclear entry, instead suggesting that the NEST device allows nuclear entry by mechanoporation. To further appreciate the onset of DNA expression between transfection methods, we calculated the relative change of DNA expression between time spans (Fig. 3d). When using the NEST device, about 50% of transfected cells expressed the protein in under 1 h, and 80% of transfected cells showed protein expression by 4 h. This performance is similar to microinjection, where most of the DNA is expressed in 1 h, but the NEST device can achieve throughputs that exceed those of microinjection. The onset of protein expression for PEI and lipid-based chemical transfections occurred between 4-12 h and the relative expression was maintained at 12-24 h. Electroporation showed the major peak of relative DNA expression at 12-24 h. These findings support the interpretation that NEST porates the nuclear envelope since it produces expression times similar to microinjection, wherein the plasmid is directly injected into the nucleus of the cell.

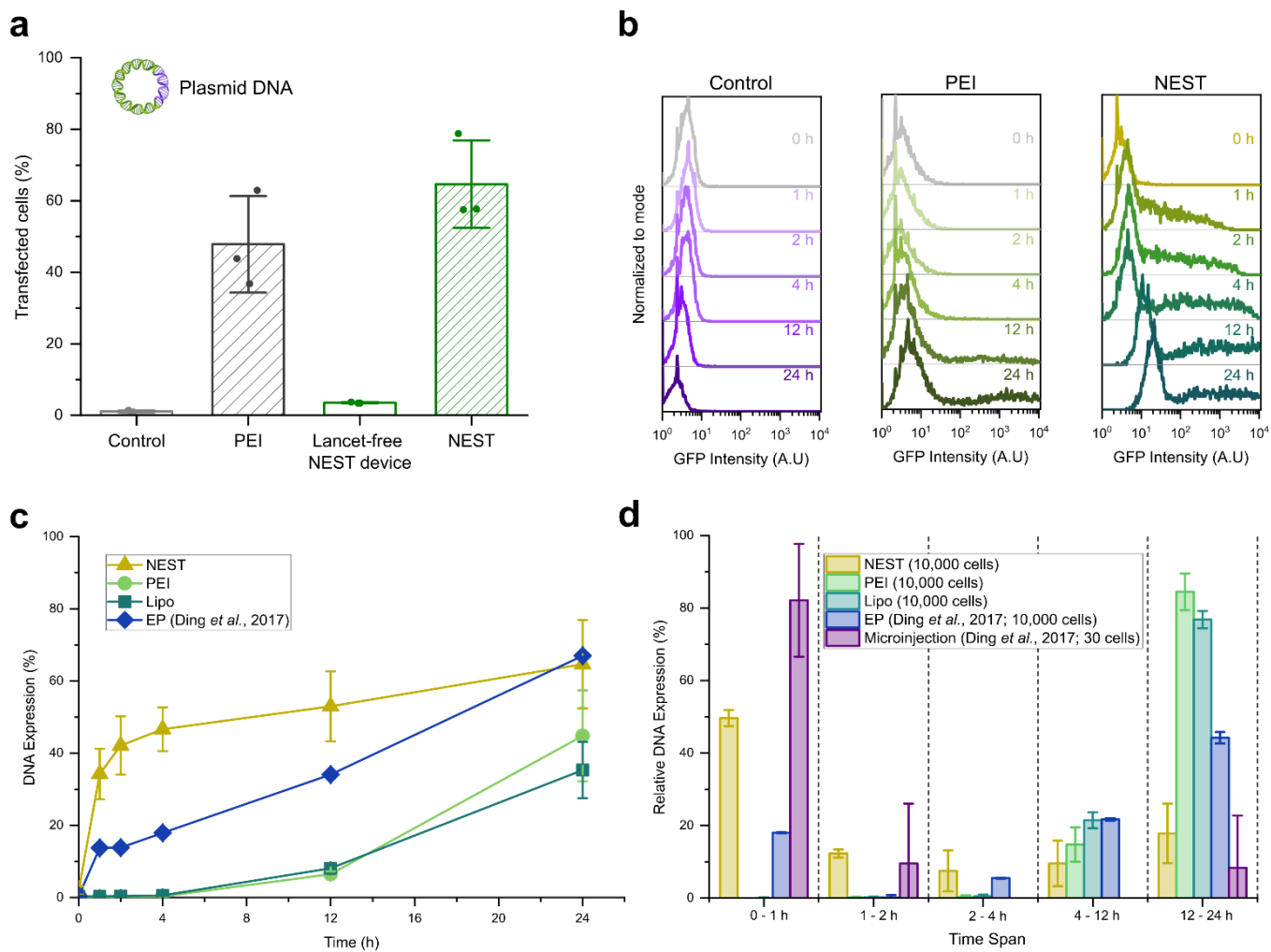

Figure 3: **Expression of EGFP-encoding plasmid DNA in HeLa cells. (a)** Comparison of EGFP-encoding plasmid DNA transfection to HeLa cells using different technologies. Transfection efficiency was evaluated via flow cytometry at 24 h after treatment. Each data point represents a sample of 10,000 cells; error bars indicate s.d. (n = 3). **(b),** Flow cytometry histogram comparison of EGFP plasmid DNA after transfection. Flow cytometry assessment was done at time points 0, 1, 2, 4, 12 and 24 h. Each histogram point represents a sample of 10,000 cells. **(c),** Quantification of plasmid DNA transfection temporal dynamics after different transfection methods. Each data point represents a sample of 10,000 cells; error bars indicate s.d. (n = 3). **(d),** Relative changes of DNA expression between time points 0, 1, 2, 4, 12, and 24 h, compared to data from a previous publication under nearly identical cell culture conditions. The relative change in DNA expression is calculated as the percentage of total final expression that manifests in each time span. For any given condition, the sum of the relative expression across all time spans is equal to 100%. The error bars indicate s.d. (n = 3).

**Cellular integrity after NEST treatment.** To assess the impact of NEST treatment on cell proliferation, we treated HeLa cells with the NEST device, both with and without pDNA delivery (Fig 4a). In all cases, including the untreated control, the cells showed an exponential growth rate, which is typical for HeLa cells. The plotted lines in Fig. 4a follow an exponential trendline with an $R^2$ greater than 0.996, demonstrating that the exponential growth rate of the cells is unaffected by NEST treatment.

To quantify the impact of NEST treatment on DNA damage, we used a commercially available fluorescence microscopy-based DNA damage analysis kit that tracks the recruitment of histones to double-stranded DNA breaks. We compared the NEST transfection system to the Lipofectamine 3000® system, which includes the reagent P3000 for enhanced transfection efficiency (Fig. 4b). All samples showed histone recruitment to double stranded DNA breaks in less than 5% of cells, with a student's T test showing no statistically significant differences between any test conditions.

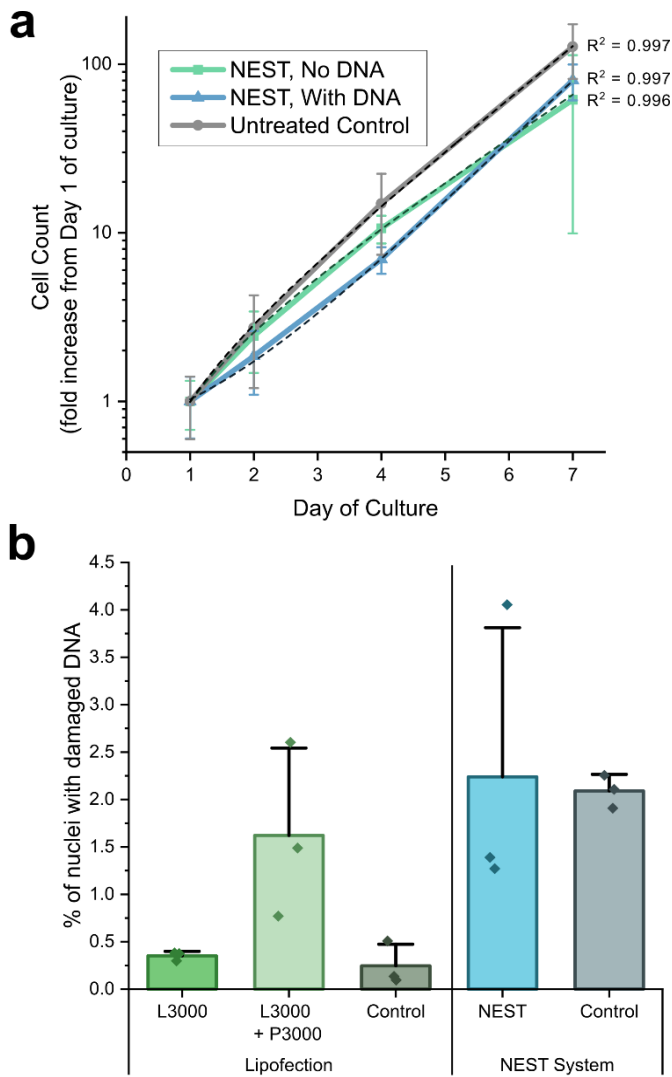

Figure 4: **HeLa cell health after NEST treatment. (a)** proliferation of cells after NEST treatment, either with EGFP-encoding pDNA or without. All three lines show an exponential growth rate (note the log scale on the Y-axis), which is expected for this cell type. **(b)** DNA damage to cells after transfection by the Lipofectamine 3000 system or the NEST system. DNA damage was assessed by phospho-histone recruitment to nuclei using a commercially available kit (see methods for details). For NEST control samples, cells were co-incubated with naked pDNA but not treated with the NEST device. Each data point represents an image area containing at least 100 nuclei. Error bars show s.d., n=3.

## Discussion

Nuclear envelope poration serves as an important strategy for rapid expression of exogenous plasmid DNA in target cells. The plasma membrane and the nuclear envelope serve as a barrier for molecular entry into the cytoplasm and nucleoplasm, respectively. During intracellular delivery, the breakdown of the nuclear envelope is necessary to allow molecules into the nucleus for a functional outcome. Nuclear envelope breakdown can result from either regulated cell division phases or physical disruption.[28] Unfortunately, cell division is a highly regulated process, and the rate of cell division strongly depends on cell types.[34–36] Thus, the ability to precisely control the disruption of the nuclear envelope is key for effective and efficient gene delivery, but balancing between sufficient pore size for efficient cargo delivery while maintaining cell recovery can be challenging, especially in high throughput.

In chemical transfection, the onset of DNA expression occurs after 12 h, which coincides with the cell division timeline and suggests that chemical transfection is dependent on cell cycle[33]. However, without a chemical vector, the half-life of exogenous DNA in the cytoplasm is 50-90 minutes due to cytosolic nucleases, resulting in the reduced expression of unprotected DNA delivered to the cytoplasm.[29] Even with a chemical vector, cytosolic DNA-vector complexes are prone to lysosomal degradation, preventing nuclear uptake during mitosis and inhibiting exogenous DNA expression[37]. The NEST device demonstrated efficient plasmid DNA transfection with rapid protein expression by nuclear envelope mechanoporation, which allowed exogenous plasmid DNA to immediately gain access to the nucleoplasmic space for subsequent transcription and translation into protein.

## Conclusions

In this study, the NEST device was able to precisely porate cell membranes and nuclear envelopes with a throughput of up to tens of millions of cells per minute, and to deliver small and large cargos to a variety of cells. The device's unique, non-viral poration method combined with its high-throughput capabilities may lend itself to application in CAR T-cell therapy, where rapid and efficient gene delivery is a necessity. Although this study did not specifically investigate the efficacy of this device in immune cell engineering, and thus this application has not been directly validated, prior studies have emphasized the need for improved delivery methods in immune cells[6]. More broadly, the applicability of this technology to diverse cell types demonstrates its potential to advance our ability to manufacture better engineered living cells that have a myriad of therapeutic applications, including in treating incurable diseases.

## Materials and methods

**Device fabrication.** The NEST device was fabricated by cleaning a 4-inch silicone wafer (El-Cat Inc., Ridgefield Park, NJ) with oxygen plasma, then spin coating the wafer with photoresist Microposit S1813 (Kayaku Advanced Materials, Inc, Westborough, MA). After UV exposure (EVG EV-420, St. Florian am Inn, Austria) and development in MF-26A developer (Kayaku Advanced Materials, Inc, Westborough, MA), the wafer underwent the DRIE Bosch process (Oxford Instruments Plasmalab 100 ICP, Concord, MA) **(Supplementary Figure 1)**. The target width of the microchannels was ranged from 8 – 20 µm, depending on the cell type to be treated. To generate sharp-tip nanostructures, silicon dioxide was grown on the etched wafer (ProTemp Diffusion Furnace, Santa Clara, CA), and then the oxide layer was etched using BOE 6:1. The inlet and outlet holes of the device were laser cut (ULS-25, Universal Laser Systems, Inc., Scottsdale, AZ) on a Borofloat 33 wafer (UniversityWafer, Inc., South Boston, MA). The microstructured silicon wafer and the lasercut Borofloat wafer were aligned and anodically bonded (EVG 520 IS, St. Florian am Inn, Austria), then diced into individual devices (DAD3220, Tokyo, Japan).

**Cell culture.** HeLa cells (HeLa S3) were a gift from the Prof. Xuedong Liu Lab at the University of Colorado Boulder. Cells were cultured in DMEM (Gibco™ 10566024, Thermo Fisher Scientific, Waltham, MA) with 10% Fetal Bovine Serum (Gibco™ A3160602, Thermo Fisher Scientific, Waltham, MA) and 1% Penicillin-Streptomycin (Gibco™ 15140122, Thermo Fisher Scientific, Waltham, MA). T-25 flasks (FB012935, Fisher Scientific, Hampton, NH) with 5ml of cell culture medium were used.

CT26 cells (ATCC CRL-2638, Manassas, VA) were cultured using the same medium and culture procedures as HeLa cells. Jurkat (Jurkat E6-1, ATCC TIB-152, Manassas, VA) were cultured in Corning RPMI 1640 medium (10040CV, Corning Inc, Corning, NY) with 10% Fetal Bovine Serum and 1% Penicillin-Streptomycin. Jurkat cells were maintained at a density between 100,000 to 1,000,000 cells per mL of culture medium. A20 cells (ATCC TIB-208, Manassas, VA) were cultured in RPMI 1640 medium (Corning® 10040CV, Corning Inc, Corning, NY) with 10% Fetal Bovine Serum, 1% Penicillin-Streptomycin, 0.05 mM 2-mercaptoethanol (M3148, Sigma-Aldrich®, St. Louis, MO), 1% MEM Nonessential Amino Acids (25025CI, Corning Inc, Corning, NY), 1% sodium pyruvate (25000CI, Corning Inc, Corning, NY) and 1% HEPES buffer (25060CI, Corning Inc, Corning, NY). A20 cells were maintained at a density between 100,000 to 1,000,000 cells per mL of culture medium.

**Fluorescent imaging of plasma membrane and nuclear envelope wounds.** HeLa CHMP4B-GFP cells, a generous gift from Dr. Martin P. Stewart, were cells were stained with Hoechst 33342 (Invitrogen™ R37605, Thermo Fisher Scientific, Waltham, MA) to observe the nucleus. The cells were harvested and resuspended in DPBS, then treated with the MemBrite™ Fix 594/615 Cell Surface Staining Kit (30096, Biotium, Inc., Fremont, CA) per the manufacturer's instructions.

After membrane staining, the cells were suspended in DPBS supplemented with 0.1% F-68 Pluronic and 2mM EDTA. The cell suspension was flowed into the NEST device using a pressure-driven syringe pump [17]. After NEST treatment, the cells were incubated with antibiotic-free warm growth medium to initiate cell recovery. Three minutes after adding medium, cells were fixed at different recovery time points using a mixture of 2% paraformaldehyde and 3% glyoxal. A 4% glyoxal solution was prepared as indicated in Richter, et al., 2018[38]. 3 mL of the glyoxal solution was mixed with 1 mL of 16% PFA in DPBS and 2 mL DPBS, producing 6 mL of a fixative with 2.67% PFA and 2% glyoxal. 100 μL of cell sample was fixed in 300 μL this solution, diluting the fixative to a final concentration of 2% PFA and 3% glyoxal, for 15 minutes at room temperature. Although further preparation steps are required after cell fixation, fixing the cells allows for these preparation steps and the imaging steps to proceed without cell recovery during this time.

Then, the cells were washed twice with DPBS and permeabilized with 0.1% triton X-100 for 15 minutes at room temperature. The nuclear envelope was stained using a lamin A/C mouse monoclonal antibody conjugated with AlexaFluor® 647 (41357, Cell Signaling Technology, Inc., Danvers, MA) diluted 1:50 in a solution of DPBS with 0.5% BSA, and incubated at room temperature for 1h. The cells were rinsed twice with DPBS and stored under refrigeration protected from light before confocal microscopy and analysis.

The fixed, treated HeLa CHMP4B-GFP cells were plated on a black-framed 96-well plate (P96-1.5P, Cellvis, Mountain View, CA) and centrifuged at 200 rcf for 2 minutes to promote settling of the suspended cells to the bottom surface. A laser scanning confocal microscope (Nikon A1, Melville, NY) equipped with a 60× oil immersion objective was used to capture images and Z-stacks of individual HeLa CHMP4B-GFP cells. The confocal fluorescence images were analyzed using Fiji ImageJ (Version 1.53c).

Analysis of confocal images revealed poration of both the cell membrane and nuclear envelope **(Fig. 1d** in the main manuscript), as evidenced by an absence of fluorescent signal in a portion of the nuclear envelope. This observation is further strengthened by an absence of fluorescent signal in a portion of the cell membrane coinciding with the location of the discontinuity in the nuclear envelope. Taken together, these observations suggest pore generation in both the plasma membrane and nuclear envelope via puncture with the sharp-tip nanostructure. These areas were taken as the location of plasma membrane and nuclear envelope pores for the purposes of measuring the diameter of these pores.

**NEST intracellular delivery and transfection.** To prepare the sample for intracellular delivery, cells were harvested and resuspended in DPBS with 0.1% F-68 Pluronic (Gibco™ 24040032, Thermo Fisher Scientific, Waltham, MA) and 2mM EDTA with a cell density of $1\times10^6$–$5\times10^6$ cells/ml. Fluorescent-labeled dextran molecules with a molecular weight of 3 kDa (Invitrogen™ D7132, Thermo Fisher Scientific, Waltham, MA), 70 kDa (Invitrogen™ D1823, Thermo Fisher Scientific, Waltham, MA), or 2000 kDa (FD2000S, Sigma-Aldrich, Saint Louis, MO) were added to the cell suspension as cargo delivery materials at a concentration of 0.1 mg/ml. The cell-cargo mixtures were then loaded into the syringe and flowed into the NEST device using a pressure-driven syringe pump.[17] After NEST treatment, the cells were collected in a 1.5 ml microcentrifuge tube and incubated on ice for 5 minutes. Subsequently, fresh warm growth medium was added to the tube, and the cells were incubated at 37°C for 15 minutes. After complete recovery, the cells were washed twice with PBS and then analyzed by multicolor flow cytometry (BD FACSCelesta™ Cell Analyzer, BD Biosciences, Franklin Lakes, NJ). Propidium iodide (P4170, Sigma-Aldrich, Saint Louis, MO) was added immediately prior to flow cytometry analysis at a concentration of 1 μg/ml as a cell viability indicator.

For imaging the nuclear localization of dextran after delivery, the cells were prepared as described above. Fluorescein-labeled 70 kDa dextran was added to the cell suspension at a concentration of 1 mg/mL. A higher concentration was used for the imaging study to enhance the brightness of the cargo and enable microscopy imaging. After recovery, the cells were fixed with 2% PFA for 15 minutes at room temperature, then rinsed three times in DPBS before confocal microscopy. The samples were stored under refrigeration and protected from light before microscopic analysis.

For naked plasmid DNA transfection, the harvested cells were suspended in OptiMEM (Gibco™ 11058021, Thermo Fisher Scientific, Waltham, MA) with 0.1% F-68 Pluronic. The gWiz-GFP plasmid DNA (Aldevron, Fargo, ND) was added to the cell suspension at a concentration of 0.1 mg/ml. After NEST treatment, the cells were collected in a microcentrifuge tube and incubated on ice for 5 minutes. Antibiotic-free warm growth medium was then added to the treated cells. The cells were incubated for 10 minutes until recovery, then washed once with PBS and cultured in a 24-well plate in growth medium. To analyze the dynamics of transfection efficiency, the cells were fixed at different time points using 2% paraformaldehyde (15714, Electron Microscopy Sciences, Hatfield, PA) for 15 minutes at room temperature. The fixed cells were washed twice with PBS and protected from light in a 7°C refrigerator until analysis in flow cytometry. Plasmid transfection was also observed with an epifluorescence inverted microscope (Nikon Eclipse Ti2 Inverted microscope, Melville, NY), and images were captured using a digital camera (Hamamatsu C11450 ORCA Flash-4.0LT, Bridgewater, NJ).

**Polyethylenimine (PEI) transfection.** To prepare for PEI transfection, HeLa cells were seeded on 12-well plates 12-24 hours in advance until 70-90% confluence. A PEI solution was made by dissolving solid PEI (23966-100, Polysciences, Inc., Warrington, PA) in deionized water with hydrochloric acid (A144S-500, Fisher Scientific, Hampton, NH) to reduce the pH below 3, then adjusting the pH to 7.5 with sodium hydroxide (S88753, AquaPhoenix Scientific, Hanover, PA) after the PEI fully dissolved. The PEI solution was then sterile filtered using a 0.22 μm syringe filter. To form the PEI-DNA complex, 1 μg plasmid DNA and 6.5 μg PEI were added to a 150 mM NaCl solution, and the mixture was incubated at room temperature for 10 minutes. For each well, the PEI-DNA complex was added to 500 μL of growth medium at different time points to observe the transfection dynamics. Cells were harvested with TrypLE Express Enzyme (Gibco™ 12604013 Thermo Fisher Scientific, Waltham, MA) and fixed with 2% paraformaldehyde (15714, Electron Microscopy Sciences, Hatfield, PA) for 15 minutes at room temperature. The fixed cells were washed twice with PBS and protected from light in a 7°C refrigerator until analysis by flow cytometry.

**Lipofectamine transfection.** To prepare for lipofectamine transfection, HeLa cells were seeded on 24-well plates 24 hours in advance until 70-90% confluence. Lipofectamine 2000 (11668019, Thermo Fisher Scientific, Waltham, MA) was used as per the manufacturer's protocol. 50 μL of Opti-MEM (Gibco™ 11058021, Thermo Fisher Scientific, Waltham, MA) was mixed with 2 μL of Lipofectamine. In parallel, 50 μL of Opti-MEM was mixed with 0.8 μg of DNA. Both Opti-MEM mixtures were combined and incubated for 20 minutes at room temperature. For each cell culture well, the Lipofectamine-DNA complex was added to 500 μL of growth medium at different time points to observe the transfection dynamics. The cells were harvested and fixed as described previously.

**Cell proliferation assessment.** Hela cells were harvested and treated with the NEST device as described previously, either with pDNA present or absent in the flow buffer. The cells were collected and plated into a 12-well plate with culture medium, then incubated at 37°C with 5% $CO_2$. Over the course of cell culture, the cells were imaged (Nikon Eclipse Ti2 Inverted microscope, Melville, NY) and each cell was manually counted. When the cells reached confluence, they were harvested with TrypLE as described previously, then split 1:8 and re-plated into a new 12-well plate. The imaging, harvesting, and replating processes continued until 7 days after initial NEST treatment.

**DNA Damage assay.** A commercially available DNA damage assay kit (Invitrogen™ H10292, Thermo Fisher Scientific, Waltham, MA) was used to assess DNA damage in NEST-treated cells. First, HeLa S3 cells were harvested and pDNA was transfected by the NEST device as described previously. After treatment, the cells were cultured for 24h before fixing and staining. Samples were prepared in triplicate.

Lipofectamine samples were prepared by plating HeLa S3 cells 48 h to 70-90% confluence prior to the addition of lipofection reagents. Lipofectamine 3000 (Invitrogen™ L3000001, Thermo Fisher Scientific, Waltham, MA) was used according the manufacturer's protocol, either including or omitting the included P3000 reagent. After the addition of the reagents and pDNA, the cells were cultured for 24h before fixing and staining. Samples were prepared in triplicate.

After 24h of transfection, the cells were fixed in 4% PFA for 15 minutes at room temperature, then rinsed with DPBS and permeabilized with a 0.1% Triton-X solution for 15 minutes. After permeabilization, the fixed cells were rinsed again and incubated in a solution of DPBS with 0.5% BSA as a blocking buffer for 1 h at room temperature. The blocking buffer was removed, and the cells were incubated in a solution of pH2AX mouse monoclonal antibody at the manufacturer's recommended concentration for 1 h. the antibody was then removed, the cells were rinsed three times with DPBS, and a solution containing a secondary antibody (Alexa Fluor® 555 goat anti-mouse IgG) and Hoechst 33342 was added to the wells at the manufacturer's recommended concentrations and incubated for 1 h at room temperature. After a final rinse in DPBS, the cells were ready for imaging.

The cells were imaged under visible and fluorescent light (Nikon Eclipse Ti2 Inverted microscope, Melville, NY). Fiji ImageJ was used to count the number of nuclei stained by Hoechst and the number of nuclei stained by the pH2AX antibody in each image. These values were used to calculate the percentage of cells showing evidence of histone recruitment to double-stranded DNA breaks, indicating DNA damage.

## Author contributions

data; L.A. wrote the paper; X.D. supervised research and revised the manuscript.

## Conflicts of interest

A patent based on this work was granted.

## Data availability

The data supporting this article have been included as part of the Supplementary Information.

## Acknowledgements

This research was supported by the NIH Maximizing Investigators' Research Award (MIRA) 1R35GM142817 and the University of Colorado Boulder Research & Innovation Seed Grant. L.A. and A.K.F. acknowledge support from the Teets Family Endowed Doctoral Fellowship. Part of the microfabrication of the devices was performed in the Utah Nanofabrication facility at the University of Utah.